# ELECTROMAGNETIC ANALYSIS OF AN ULTRA-LIGHTWEIGHT CIPHER: PRESENT

Nilupulee A. Gunathilake[1], Ahmed Al-Dubai[2],
William J. Buchanan[1], Owen Lo[1]

[1]Blockpass ID Lab, School of Computing, Edinburgh Napier University, UK
[2]School of Computing, Edinburgh Napier University, UK

## ABSTRACT

*Side-channel attacks are an unpredictable risk factor in cryptography. Therefore, continuous observations of physical leakages are essential to minimise vulnerabilities associated with cryptographic functions. Lightweight cryptography is a novel approach in progress towards internet-of-things (IoT) security. Thus, it would provide sufficient data and privacy protection in such a constrained ecosystem. IoT devices are resource-limited in terms of data rates (in kbps), power maintainability (battery) as well as hardware and software footprints (physical size, internal memory, RAM/ROM). Due to the difficulty in handling conventional cryptographic algorithms, lightweight ciphers consist of small key sizes, block sizes and few operational rounds. Unlike in the past, affordability to perform side-channel attacks using inexpensive electronic circuitries is becoming a reality. Hence, cryptanalysis of physical leakage in these emerging ciphers is crucial. Among existing studies, power analysis seems to have enough attention in research, whereas other aspects such as electromagnetic, timing, cache and optical attacks continue to be appropriately evaluated to play a role in forensic analysis.*

*As a result, we started analysing electromagnetic emission leakage of an ultra-lightweight block cipher, PRESENT. According to the literature, PRESENT promises to be adequate for IoT devices, and there still seems not to exist any work regarding correlation electromagnetic analysis (CEMA) of it. Firstly, we conducted simple electromagnetic analysis in both time and frequency domains and then proceeded towards CEMA attack modelling. This paper provides a summary of the related literature (IoT, lightweight cryptography, side-channel attacks and EMA), our methodology, current outcomes and future plans for the optimised results.*

## KEYWORDS

*Side-channel attacks, electromagnetic analysis, lightweight cryptography, PRESENT and IoT.*

## 1. INTRODUCTION

Internet of things (IoT) is a wide-spread infrastructure that consists of millions of connected devices. The main purpose of the IoT is to produce useful insights through data analytics. This causes collection, process and distribution of information continuously privately and publicly. Because of that, avoidance of data breaches is challenging. A summary of overall IoT communication strategies, technologies and challenges is accessible in [1] and [2]. IoT devices are constrained in terms of physical size, memory allocation and data rates (kbps) because they target sensor-based applications, *i.e., wearables to monitor health features, fault detection in factory automation, vehicular communications with real-time insights, etc.* Therefore, conventional cryptographic algorithms such as Advanced Encryption Standard (AES) used in general computing devices are impractical to function well on the IoT platform, due to long block sizes and key lengths that would require high-end processors. As a solution, a concept towards light versions of the ciphers named lightweight cryptography was introduced [3 - 4].

One of our recent publications [5] covers a complete survey of lightweight cryptography. It includes history, categories of all existing lightweight ciphers (symmetric, asymmetric and hash) up to today, along with their parametric values. Furthermore, it critically analyses cryptologic as well as cryptanalysis studies available in the literature. Among hundreds of recommended lightweight cryptographic algorithms and their versions, some of the well-known ciphers in the field are included in Table 1 compared to the AES.

Table 1. Lightweight ciphers versus AES

| Cipher | Architecture | Block Size (b) | Key Size (b) | Rounds | S-box (b) |
|--------|--------------|----------------|--------------|--------|-----------|
| AES | SPN | 128 | 128/192/256 | 10/12/14 | 8 |
| KLEIN | SPN | 64 | 64/80/96 | 12/16/20 | 4 |
| PRESENT | SPN | 64 | 80/128 | 31 | 4 |
| Fantomas | LS-design | 128 | 128 | 12 | 5 |
| LBlock | Feistel Network | 64 | 80 | 32 | 4 |
| Piccolo | Feistel Network | 64 | 80/128 | 25/31 | 4 |
| PRINCE | FX construction | 64 | 128 | 12 | 4 |
| Simon | Feistel Network | 64 | 96/128 | 42 | - |
| Speck | Feistel Network | 64 | 96 | 26 | - |
| LED | SPN | 64 | 64/128 | 32/48 | 4 |
| SPN: Substitution Permutation Network, LS: Linear diffusion and non-linear S-box | | | | | |

Side-channel attacks are external phenomena that create an environment for unauthorised third parties to investigate sensitive information leakage through physical parameters such as power consumption, thermal radiation, time duration, cache files and optical changes when cryptographic functions are running on hardware. An attacker can approach the tasks via invasive and non-invasive ways. Invasive methods involve de-packaging the chip to get the connectivity to its inside elements, *i.e., access to a data bus to monitor data transfers, etc.*, whereas non-invasive trials investigate externally available details only, *i.e., energy drainage, electromagnetic (EM) emission, etc*. The major areas in side-channel attacks are, but not limited to:

- **Probing attack**: Direct observations of the internal parameters of the device.
- **Cache attack**: Cache access is monitored in a shared physical system [6, 7], *i.e., virtual machines, cloud, etc.*
- **Data remanence attack**: Sensitive data is recycled after deletion, *i.e., coldboot attack [8].*
- **Timing attack**: Running time is monitored.
- **Acoustic attack**: Sound generated is concerned [9].
- **Optical attack**: The surrounding of the device is visualised to see any indication using high resolution cameras [10].
- **Fault analysis attack**: Clock, temperature, voltage, radiation, light and Eddy current[1] are measured.
- **Power analysis attack**: Simple, differential and correlation power analysis (SPA, DPA, CPA) [11 - 12].
- **EM analysis attack**: Measures of EM field around the device.

---

[1] Loops of electrical current induced within conductors by a changing magnetic field in the conductor according to Faraday's law of induction

Preliminary analysis of those scenarios is essential to prevent physical leakages for guaranteed security of a cipher. General countermeasures that could be taken to mitigate weaknesses in physical security are:

- Cryptographic operation obfuscating enabled firmware [13].
- Randomisation of the operation sequences and or lookup tables in ciphers.
- Access of critical data using pointers instead of values when data structures are chosen.
- Asynchronism of the clock in the chip with respect to the cryptographic functions.
- Design of mathematical models in a manner that the leakage is misguiding.
- Application of masking technique where appropriate (research [14] and [15] prove that masking would be insufficient against EM attacks).
- EM shielding via suitable materials when the chip/item is manufactured, *i.e., inclusion of Faraday cages, etc.*
- Excessive noise addition to hiding leakages in EM radiation.

Since lightweight cryptography is still emerging in academia, the main focus can be seen in algorithm optimisation in cryptanalysis. Among the few studies, power analysis takes the highest percentage, *i.e., correlation power analysis (CPA) of PRESENT [12], Fantomas, LBlock, Piccolo, PRINCE, Simon and Speck [16], differential power analysis (DPA) of PRESENT [17], Simon and LED [18]*. In EM analysis (EMA), the study [19] shows their results for a differential EMA (DEMA) of PRESENT, the studies [20] and [21] of a correlation EMA (CEMA) of PRINCE and Twine respectively. Yet any research outcome regarding CEMA of PRESENT seems to be unavailable in the literature. Other types of attacks on these novel ciphers remain to be considered in academia as well.

## 1.1. Our Contribution

In comparison, CEMA tends to offer more efficacy than DEMA theoretically. However, results may vary practically. Due to the fact that the unavailability of a study on CEMA of PRESENT, which is the most promising lightweight block cipher to include in IoT in the near future, we thought of filling a research gap by conducting a CEMA of PRESENT against firmware robustness. Here, performances from simple analysis in both time (SEMA) and frequency (SFEMA) domains to correlation white-box attack modelling were analysed. The work is still on-going for improved results and evaluation.

The major contents of this paper are:

- A comprehensive overview about EMA including types of attack models and related mathematical equations.
- In-depth details about our attack model including code snippets, so it would help freshers in the field to begin with their own experimental setups.
- Current results of our SEMA, SFEMA and CEMA attack models.
- Discussion over the observations and real-world scenarios.
- Up-coming plans for optimising the attack model.

## 2. ELECTROMAGNETIC ANALYSIS OF PRESENT

### 2.1. Electromagnetic Analysis

Although EMA seemed to be conducted in the time domain in the past, an increased interest can be seen in the frequency domain recently. However, methods need to be well designed accordingly as just domain conversion via fast Fourier transform (FFT) would not create an accurate attack model. Our literature findings conclude an existing model classification as in Fig.1. There are two mathematical approaches used in these analyses based on Hamming calculations:

- **Hamming distance (HD) method**: XOR operation between two binary values. In a binary number, bit by bit operation is reflected here.

$$W = HD(D, R) = a.HW(D \oplus R) + b \qquad (1)$$

*where,*
*W - Hypothesised value*
*D - Intermediate value*
*R - Reference state value*
*a - Gain*
*b – noise*

- **Hamming weight (HW) method**: Non-zero elements of a binary number where the reference state is zero. For example, if a number is 10010101, then the Hamming weight would be 4.
-

$$W = a.HW(D) + b \qquad (2)$$

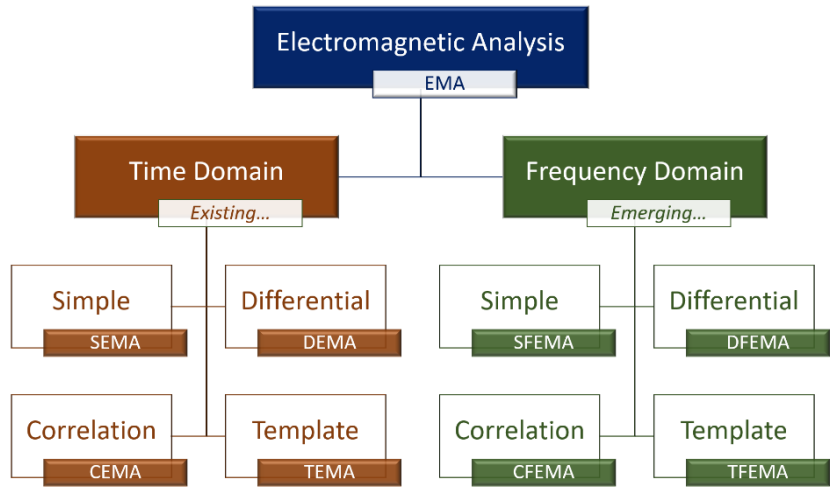

Figure 1.  General EMA classification

**SEMA and SFEMA** are visual inspections of EM traces recorded using an oscilloscope or a software-defined radio (SDR). It helps identify where the leakage occurs, but it does not involve extracting secret information such as the encryption key to breaking into the secret data. On the other hand, locating the exact operation of the cipher and guessing the key via clock information followed by brute-force analysis are possible sometimes, if thorough knowledge of the device is there [22].

**DEMA and DFEMA** [19, 23] are complex statistical models which do not require much information about the device. These derived from the differential power analysis (DPA) method introduced by [24].The analysis continues with guesses while corresponding traces are divided into two groups based on the guess for a bit to be 0 or 1. Then all traces of each group added together and averaged. Finally, the difference between the averages of group 1 and group 0 is calculated. If the trace alignment is correct, a considerable amount of spikes will illustrate supportive information for the derivation of the key.

**CEMA and CFEMA** [20, 21, 25] are efficient versions of DEMA and DFEMA respectively where grouping is not required. This also does not need knowledge of the device. The model focuses on several bits at a time. The analysis is based on the correlation of either the HD or HW method. It offers a hypothetical intermediate value that would indicate the possibility of the attack. The correlation between the EM emanation and hypothesised intermediate value could be calculated using the equation 3.

$$\rho = \frac{Cov(X,Y)}{\sqrt{Var(X).Var(Y)}} \qquad (3)$$

*where,*
$Cov(X,Y) = E\big((X - E(X)).(Y - E(Y))\big)$
$Var(X) = E(X^2) - (E(X))^2$
$Var(Y) = E(Y^2) - (E(Y))^2$
E(X) - Mean of X
E(Y) - Mean of Y

**TEMA and TFEMA** [15] require a complete copy of the device with full control. Then via pre-processing using a large number of EM traces, a template is created. It further needs capturing a small number of traces from the victim's side to complete the attack.

## 2.2. PRESENT Block Cipher

PRESENT is a lightweight cryptographic algorithm that is suitable in ultra-lightweight[2] conditions too. It is a block cipher designed by the authors of [26] in 2007. It has been approved by several international standardisation authorities, the International Organisation for Standardisation (ISO), the International Electrotechnical Commission (IEC) and the National Institute of Standards and Technology (NIST).

It is based on a SPN architecture with 64b block, 80b key and 31 rounds. It operates as in Fig. 2, offering moderate security. It targets hardware optimisation by having a tiny footprint of 1570 gate equivalent (GE) and a low power consumption of 5μW over 32 clock cycles. It has the smallest substitution box (S-box), which is 4b-4b equivalent to 28 GE as well as the simplest permutation (pLayer) resulting in 0 GE. The S-box mapping as in Table 2. Moreover, this cipher has another version with a 128b key that would require 1886 GE. The authors mention that the PRESENT is more prone to side-channel and invasive hardware attacks.

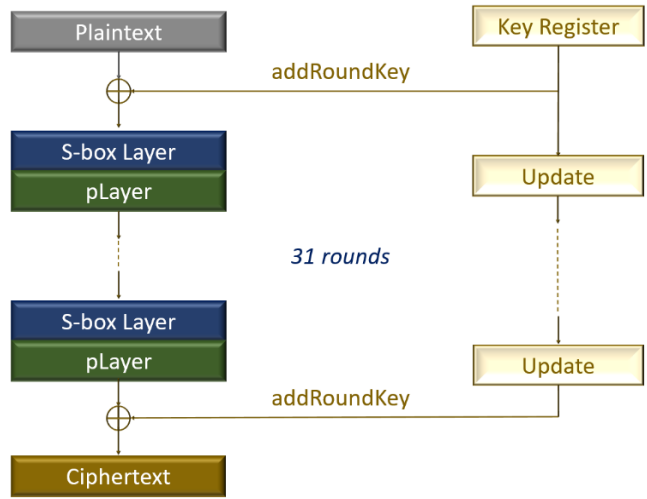

Figure 2. Operational process of PRESENT cipher

Table 2. S-box mapping of PRESENT (in hexadecimal)

| *x* | 0 | 1 | 2 | 3 | 4 | 5 | 6 | 7 | 8 | 9 | A | B | C | D | E | F |
|-----|---|---|---|---|---|---|---|---|---|---|---|---|---|---|---|---|
| *S(x)* | C | 5 | 6 | B | 9 | 0 | A | D | 3 | E | F | 8 | 4 | 7 | 1 | 2 |

In S-box, the output of the addRound key step, which is a 64b block is split into sixteen 4b blocks. Then each is fetched through the S-box layer simultaneously to look up against the mapping values. After, the block value is replaced by the resultant values.

---

[2]Correspondence only to specific areas of the algorithm, *i.e., selective micro-controllers (µC), selected cipher sections, etc.*

## 2.3. Methodology

Near field (NF) EM compatibility (EMC) probes are used to capture EM radiation around an embedded device. According to Faraday's law, changes of the magnetic flux in a magnetic field generate a voltage in the probe's loop (equation 4). Induced voltage fluctuations in the electric signal would give an idea about the EM emanation of a device. Besides, probes with smaller loops have higher frequency resolution, but less sensitive in the acquisition. The waveforms can be analysed using an oscilloscope or a software-defined radio (SDR). Here, the 80b key version of PRESENT was chosen due to the expectation of application in IoT devices. The resources used for the experimental setup are in Table 3.

(4)

$$V = 2\pi BA$$

*where,*
V - Voltage
π –Pi constant, 22/7
B - Average magnetic field
A - Area perpendicular to the magnetic field

Table 3. Resources used for the testbed

| Hardware | |
|---|---|
| Oscilloscope | KEYSIGHT InfiniiVision MSOX4101A (1GHz, 5GSa/s) |
| EM emanation acquisition | TekBox EMC NF probe set – H20, H10, H5, E5 (9kHz – 6GHz) |
| Trigger connectivity | KEYSIGHT passive probe N2894A 700MHz |
| Pre-amplifier | TekBox 40dB wide-band amplifier |
| Encryption device | Arduino UNO R3 |
| Instrument control | Microsoft Surface Laptop with an i5 processor |
| **Software** | |
| Encryption | Arduino IDE |
| Data processing | Matlab R2020b |

The hardware connectivity of the testbed is as in Fig.3. The probe was initially placed on the top of the chip in a manner that the angle becomes $90^0$ to the chip, because a study [25] mentions that it is the optimum position to acquire the highest EM radiation. However, different positions and angles are expected to take into consideration in the future. A trigger signal was used to locate the S-box function (Fig.4). The trigger function was monitored using a separate channel of the oscilloscope. Consequently, the maximum possible sampling rate became 2.5GSa/s for EM waveforms. The LED of the Arduino UNO board (pin 13) was connected to the trigger channel, and the encryption code was made in a way that the pin gets high (LED on) when the operation starts and then gets low (LED off) after the completion. In addition, primary level precautions were taken to reduce ambient and system noises such as:

- Waveforms were captured in average mode by using five encryption cycles per trace.
- Aluminium foil was used to cover the setup as an EM shield.
- The computer was operated in flight mode.

The accuracy of trace capturing and reconstruction using Matlab was verified by known test data values. The Arduino code for encryption was derived from [27] and had been previously validated using test vectors for [12].

### 2.3.1. Attack Modelling

Our model targets the S-box implementation because a successful attempt would cause key extraction straightaway. Due to its non-linear behaviour, significant differences may appear in correlation calculations. The encryption key used was AC DE FB 21 F9 23 43 75 C0 E6. The code snippet for the output implementation of the S-box in Arduino is shown below.

```
for (int i = 0; i < 8; i++)
        {
        state [i] = sBox [state [i] >> 4] << 4 | sBox [state [i] & 0xF];
        }
```

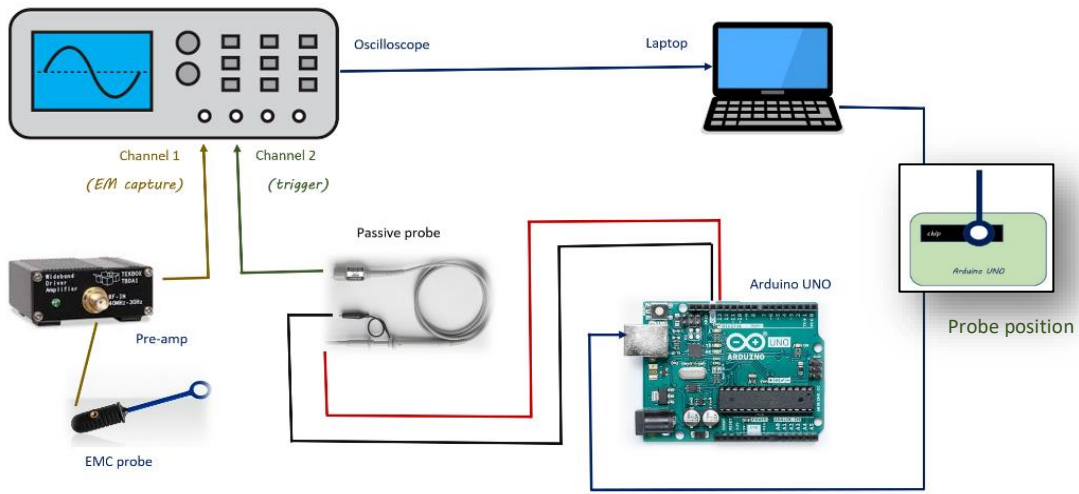

Figure 3. Hardware connectivity of the experimental setup

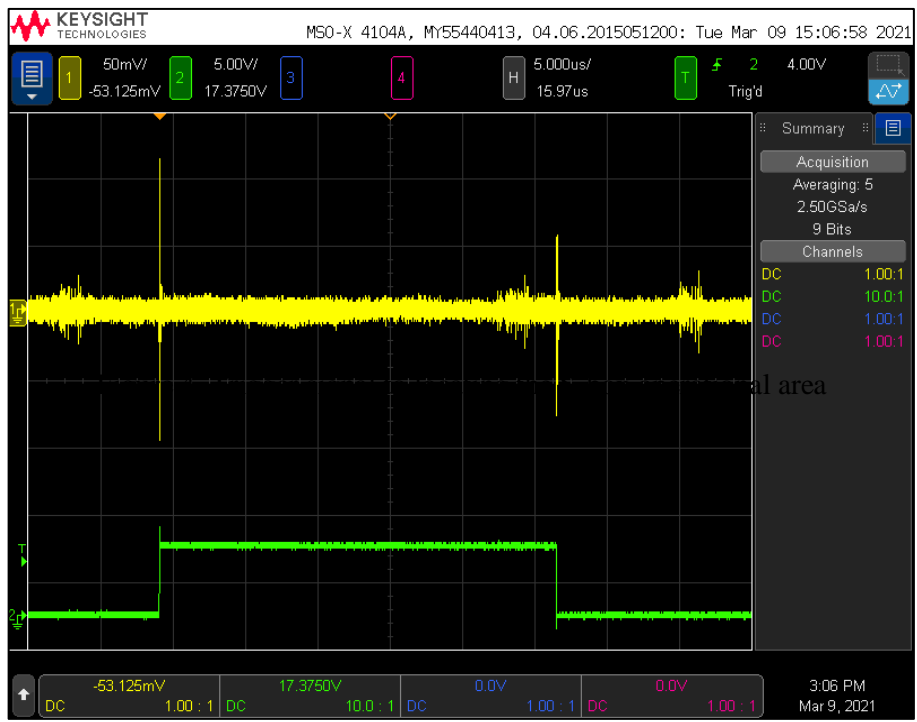

Figure 4. Trigger signal to locating the S-box operational area

The first part of the state [i], sBox [state [i] >> 4] << 4 looks up against the first four most significant bits (MSB) where sBox [state [i] & 0xF] does against the least significant bits (LSB). Next, the bitwise OR logic operator, | combines the two 4b numbers into a single 1B value. The states from 0-7 are the outputs after each S-box substitution.

The plaintext values start to increment from 0-255 in each bit (00 00 00 00 00 00 00 00 – FF FF FF FF FF FF FF FF) once a command is received via Arduino UNO's serial port from a Matlab script. Then, each waveform for each plaintext was captured and saved for post-processing. Thus, the total number of traces collected was 256, and each trace originally contained 16000 data points. When the signal was trimmed later on Matlab by extracting related values into a new matrix to filter the S-box functional area, the useable number of data points were reduced to 8800.

In CEMA modelling, the choice of method was HW because of its efficacy. The accuracy of the model was validated using the test results for [8]. The steps of the procedure:

- Firstly, hypothesised values considering both key and plaintext values from 0-255 were calculated using HW.
- Meanwhile, the actual EM values that had been recorded in 256 rows (0-256 plaintexts) for each data point (total of 8800) were compared to its correspondent hypothesised array of 256 items for each key to get the correlation coefficient values ($\rho$) using the equation 3.
- Finally, the maximum $\rho$ value was taken as the result for each data point.

The highest value in $\rho$ throughout all data points should indicate possible leakages of the key Bytes. The pseudo-code for the model is shown below.

```
for k = 0:255 (Key values),
        for p = 0:255 (Plaintext values),
                set output of AddRound key step as the input to the S-box
                lookup MSB of the input (4b)
                lookup LSB of the input (4b)
                combination of the MSB and LSB (1B)
                HW calculation and saving
        For x = 1:length of data (8800 data points for 256 waveforms),
                ρ calculation between arrays of actual values and HW array
                        if empty,
                                save key value
                                save ρ value
                        else if ρ > previous one,
                                save key value
                                update ρ value
        plot data points vs. ρ
```

## 3. RESULTS

Currently, the results are available only for the H20 and H10 probes' measurements.

### 3.1. SEMA

Fig.5 and Fig.6 show EM emission in the time domain (time versus voltage)for the S-box function for the H20 and H10 probes, respectively. Fig.7 and Fig.8 show data distribution over the operation in histogram plots for each case. Subplots in each figure illustrate the difference between encryption and non-encryption scenarios.

### 3.2. SFEMA

Fig.9 and Fig.10 show EM emission in the frequency domain (frequency versus amplitude) for the S-box function for the H20 and H10 probes, respectively. Fig.11 and Fig.12 show changes in frequency strengths over the operation in spectrogram plots for each case. Subplots in each figure visualise the difference between encryption and non-encryption scenarios.

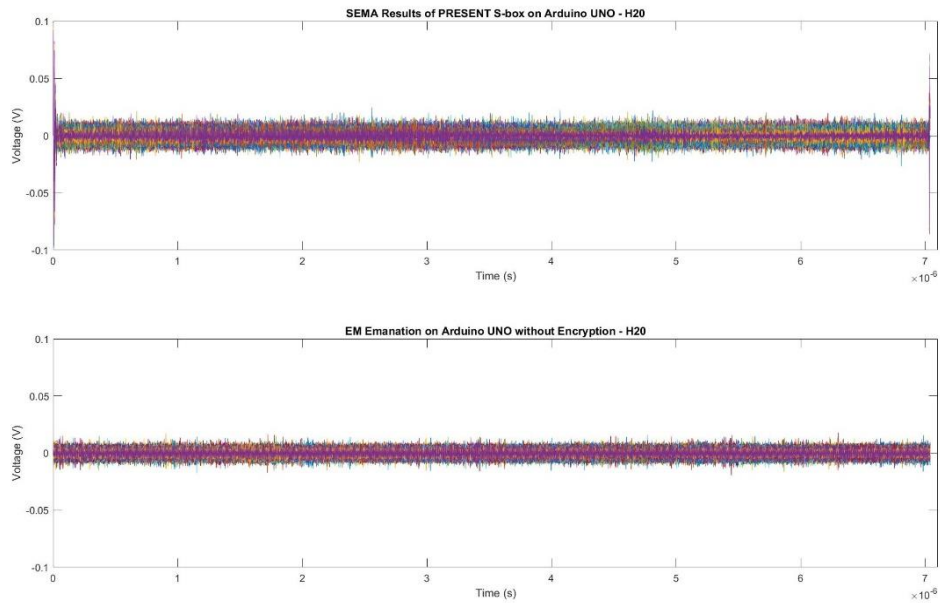

Figure 5. EM emanation in the time domain - H20 probe

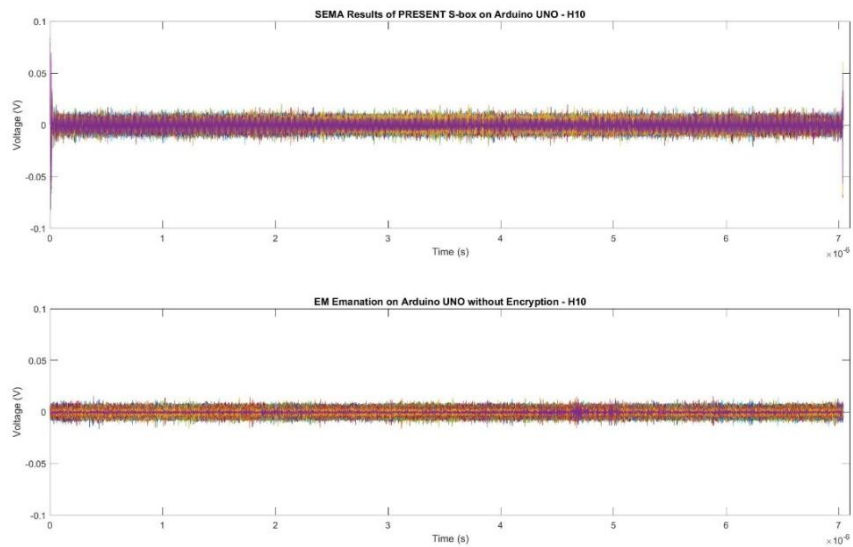

Figure 6. EM emanation in the time domain - H10 probe

## 3.3. CEMA

Fig.13 and Fig.14 illustrate the outcomes of the CEMA attacks performed using the H20 and H10 probes, respectively. Guesses for key Bytes were made by checking on notable peaks and troughs of the graph. Correspondent results for each setup are summarised in Table 4 and Table 5. Due to the repetition values in the troughs in both cases, non-encryption data values were fetched through the model to check on possible false-positive appearances because of the system noise. The results are as in Table 6 and Fig.15.

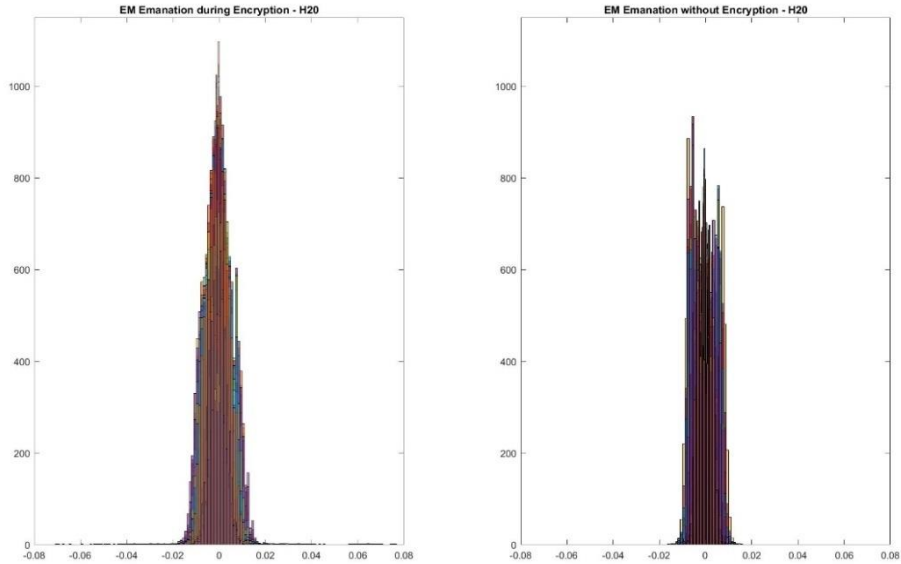

Figure 7. Histogram of the H20 probe data for the encryption and non-encryption states

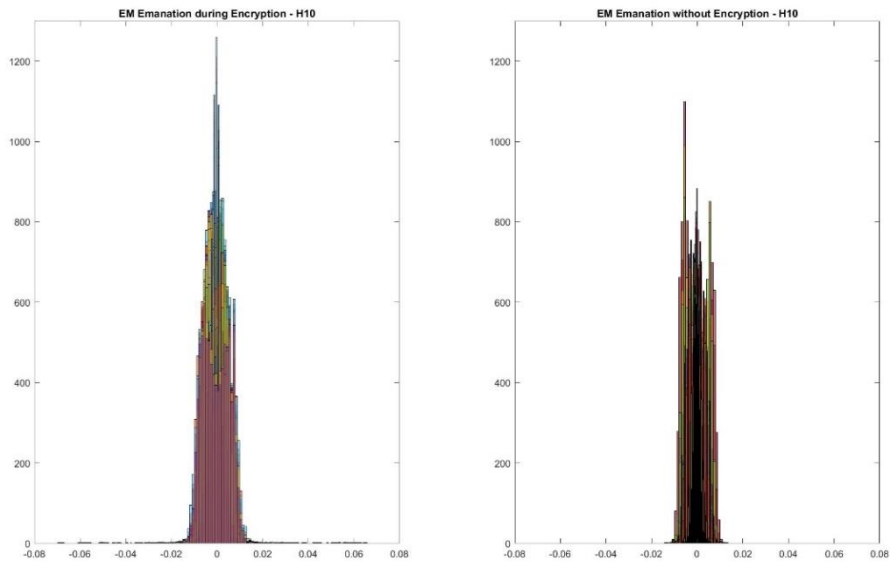

Figure 8. Histogram of the H10 probe data for the encryption and non-encryption states

Table 4. Key guess for CEMA - H20 probe

| Type | No. | Key Byte Guesses | | | | | | | | | | | |
|---|---|---|---|---|---|---|---|---|---|---|---|---|---|
| Peaks | 12 | C0 | B2 | 9E | 40 | 1A | 02 | 90 | 4F | D2 | 10 | 4D | D6 | - |
| Troughs | 13 | 90 | 99 | 9C | 90 | 90 | 99 | 99 | 90 | 90 | 90 | 99 | 91 | 72 |

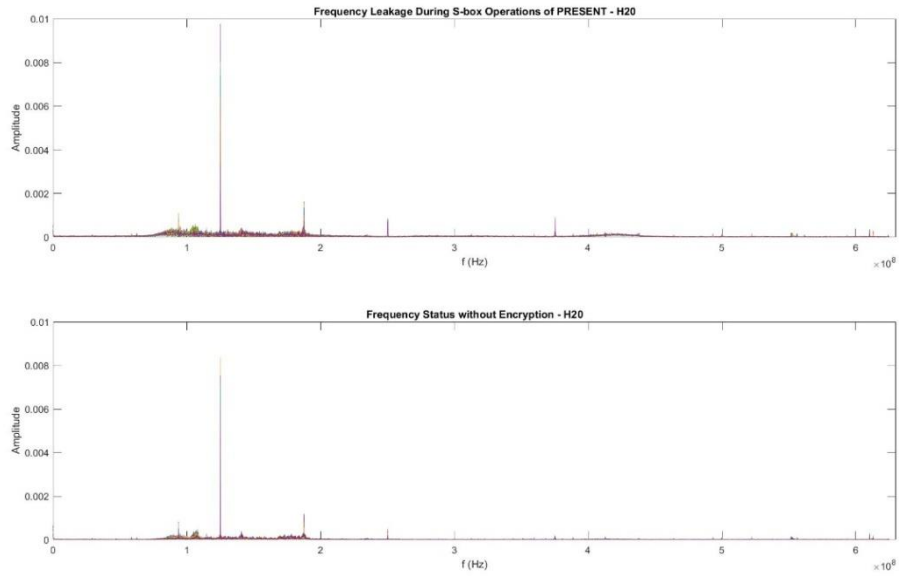

Figure 9. EM emanation in the frequency domain - H20 probe

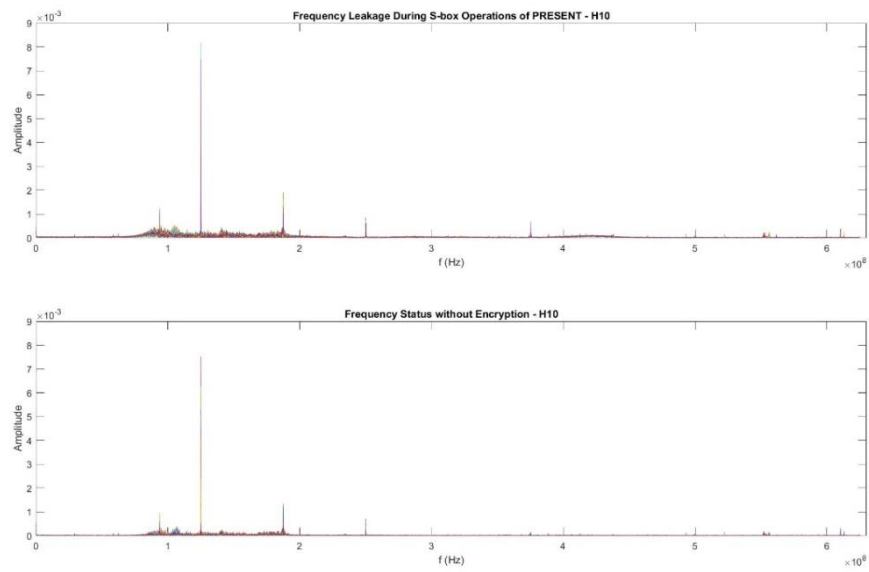

Figure 10. EM emanation in the frequency domain - H10 probe

Table 5. Key guess for CEMA – H10 probe

| Type | No. | Key Byte Guesses | | | | | | | | | |
|------|-----|----|----|----|----|----|----|----|----|----|----|----|
| Peaks | 10 | E4 | DF | 88 | 48 | C4 | D8 | E6 | 49 | 54 | 44 | - |
| Troughs | 11 | 56 | 56 | 56 | E5 | 56 | 35 | 56 | 56 | F6 | 51 | 56 |

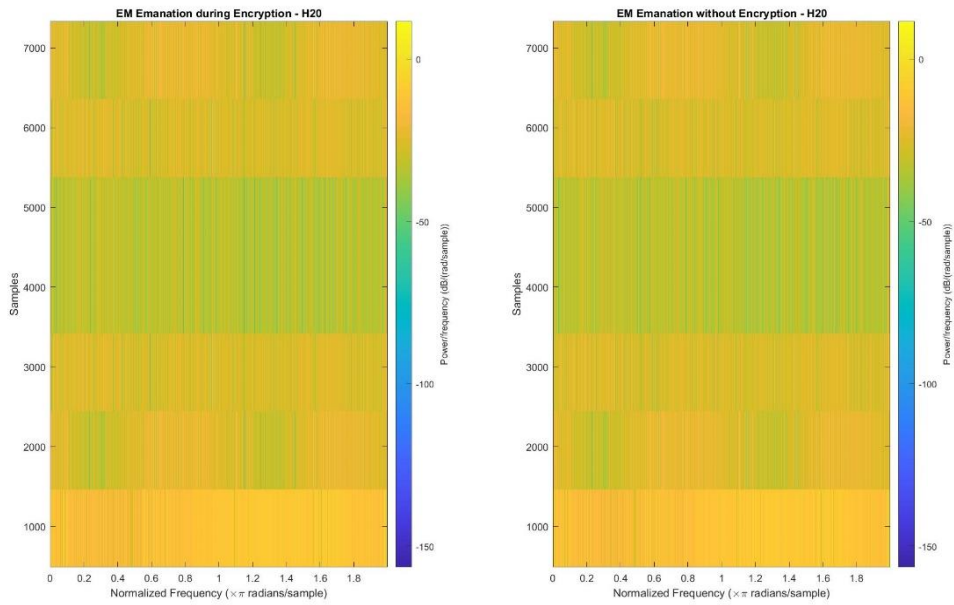

Figure 11. Spectrogram of the H20 probe data for the encryption and non-encryption states

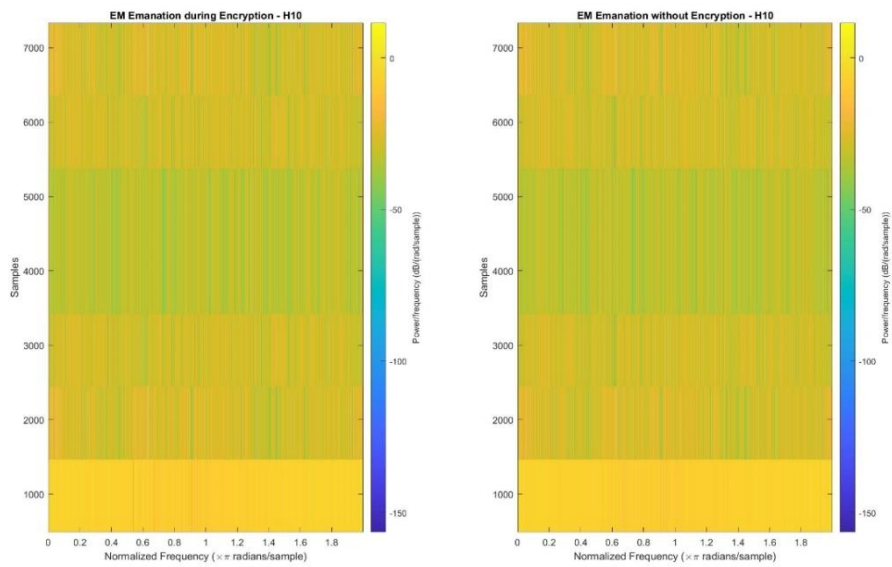

Figure 12. Spectrogram of the H20 probe data for the encryption and non-encryption states

Table 6. False positive appearance check

| Type | No. | Pattern of Repetition | | | | | | | | | | |
|------|-----|----|----|----|----|----|----|----|----|----|----|----|
| **H20** | | | | | | | | | | | | |
| Peaks | - | - | - | - | - | - | - | - | - | - | - | - |
| Troughs | 12 | C1 | C9 | CD | CD | CD | C3 | C3 | CD | C3 | C3 | C3 | CD |
| **H10** | | | | | | | | | | | | |
| Peaks | 7 | 1A | 2D | 2B | CA | 99 | 2B | 49 | - | - | - | - | - |
| Troughs | 12 | 11 | 1F | C5 | 3A | C6 | 58 | 3A | AE | 3D | 31 | C9 | C9 |

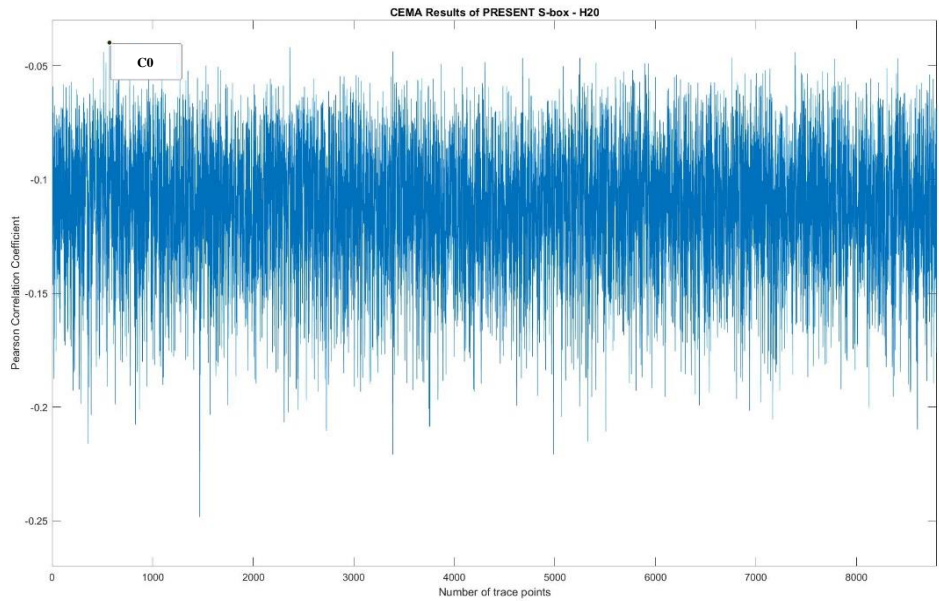

Figure 13. CEMA results for the H20 probe

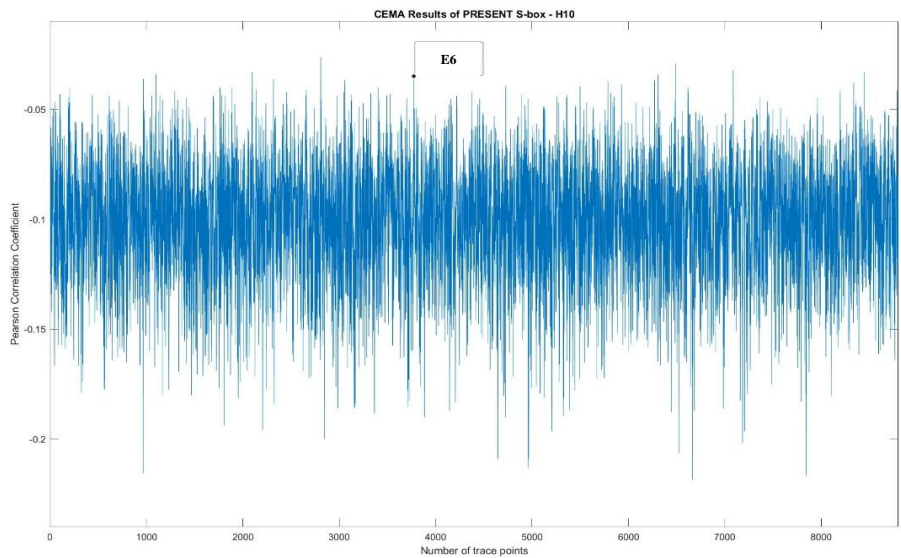

Figure 14. CEMA results for the H10 probe

## 4. DISCUSSION

Even though visible changes do not appear in the time domain graphs, slight differences can be seen in frequencies between encryption and non-encryption states. The observations of the SEMA and SFEMA are as follows:

- According to Fig.9 and Fig.10, higher density can be seen from 75MHz – 200MHz during encryption. Furthermore, a set of new frequency components seem to appear from 375MHz – 450MHz. Increased amplitude is seen in already existing components in 93MHz, 187MHz, 250MHz and 375MHz itself.

- Regarding data distribution (Fig.7 and Fig.8), values tend to shrink sharply towards zero with one peak head during encryption in both setups. Nevertheless, non-encryption state shows flattened distribution with more than one high peak.
- The outputs show similar results for both H20 and H10 probes in comparison of frequencies as well as of histograms.
- However, spectrogram diagrams do not show any significant difference between encryption and non-encryption states, but they do between the probes where the H20 probe's data has a higher density than the H10 probe's.

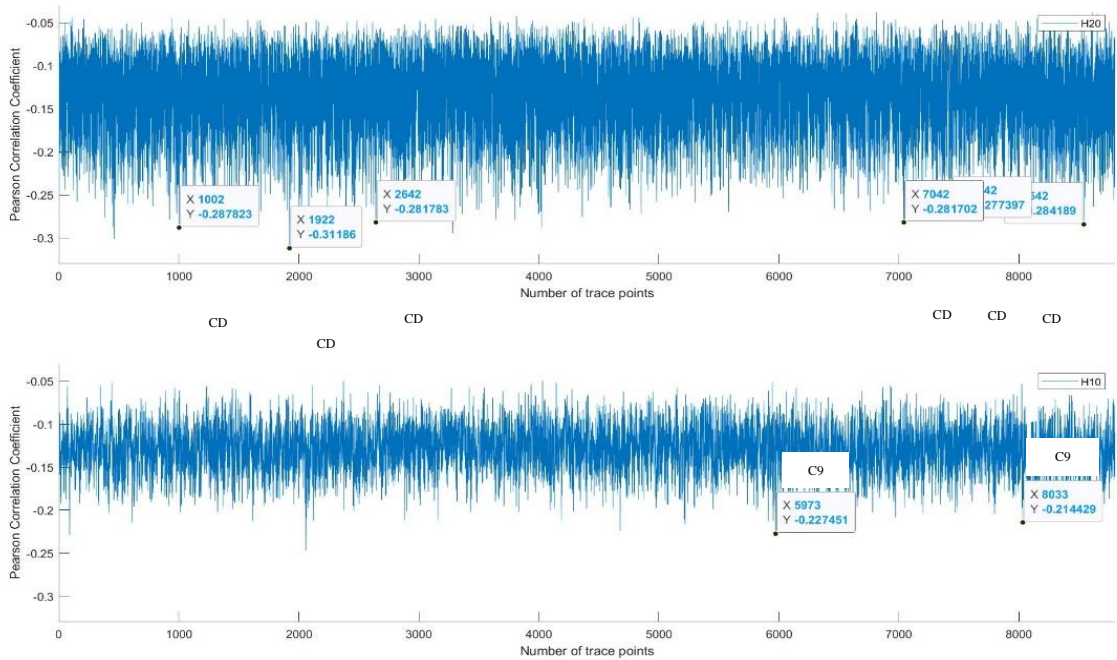

Fig.15: CEMA on non-encryption data

In spite of these differences, the induced values seem to vary on a very small scale, usually from (-20) mV to +20mV. The largest value can be seen when the trigger rises and falls, which is around 200mV.

In CEMA, the general outcomes are extremely noisy and do not appear sharp in comparison to CPA results of the study [8]. Therefore, it is quite disappointing from the attacker's side. Despite this, the H10 probe seems to offer slightly clearer results. In any case, EM emission is highly interfered with by radio waves, UHF/VHF signals, Wi-Fi transmission, Bluetooth communications and any type of EM radiations emitted by electronic devices. That is why EMA has become quite challenging and requires careful attention to perform successfully. On the other hand, it can be seen that the emission dramatically increases when the LED goes to the on position from off as well as the off position from on. Thus, it proves the probes are picking up the emanation created by the circuit board properly, but the radiation created in response to cryptographic functions may not be significant. If that is really the case, lightweight ciphers that come with such shielding would be less prone to EM side-channel attacks. Nevertheless, further analyses are needed with many more sample data to confirm the fact.

In the results, the H20 probe was able to give the 9th Byte of the encryption key in one of its peaks. Moreover, the H10 probe gave the 10th Byte correctly. For both cases, the peaks of the graph caused the correct guess rather than troughs. Anyway, verification is expected to be obtained via further tests because there is a high probability of false-positives due to the excessive noise factors mentioned in the previous paragraph. The issue is further confirmed by the repeated values in several troughs, both during the encryption and non-encryption states. This may occur due to the clock cycles running inside the chip after power-up. In contrast, repetition did not occur in the peaks. Hence, there is a greater probability of key leakages in peaks than in troughs in this model.

What is more, in the work [13], [14] and [15], more than a thousand traces have been used for their EMA research in lightweight ciphers, but in our model, only 256 waveforms for 256 plaintexts were used. Therefore, we assume that a larger number of EM waveforms would increase the probability of attacking in comparison to the quantity of power traces. The main disadvantage was the decrease in the number data points once the signal was trimmed to filter the S-box functional area. It reduced 7200 points, thus it could have been a negative impact on the outcome. To mitigate the identified weaknesses and enhance the performance of the testbed, we expect to be concern with:

- Increase in the number of waveform collections beyond a thousand for increased probability in successful attacks.
- Confirmation of correct waveform alignment to the S-box starting point.
- Bandpass filter application for identified frequency leakage ranges to reduce system noise impacts.
- Analysis of the correlation coefficient results in the frequency domain (CFEMA).
- Checking the behaviour of EM fluctuations as an electric current response rather than the voltage in the time domain.
- Finding the optimised position for probe placement on the chip.
- Choice of the ideal probe for the testbed among the five probes.
- Application of a Faraday cage using Faraday fabrics (low-cost approach compared to expensive manufactured cages) to reduce ambient EM noise.
- Comparison of the performance using different hardware, *i.e., FPGA* (this would take our research limitation beyond the firmware consideration).

While side-channel attack research is encouraged in practical cryptanalysis, it is also important to contemplate real-world scenarios. Usually, side-channel analyses are based on a particular operation of the cipher for a chosen round. In reality, encryption runs through all rounds in pipeline processing. Regarding this project's target of the S-box of PRESENT, AddRound key and pLayer operate before and after in all 31 rounds. Therefore, it would be problematic to split waveforms having proper alignment for the correct function and round. Also, the most probable attack type will be a black-box one where the attacker has no access to the model's parameters. The problem could be worse for the attacker if the manufacturer has properly taken countermeasures against possible EM leakages. Furthermore, complete noise reduction of either ambient or system is an unavoidable task.

## 5. CONCLUSIONS

In this 5G era, connected devices are increasing massively. The IoT has become a widespread infrastructure in communications and analytics of data. However, provision of sufficient security is challenging because of the resource-constrained environment in the IoT platform. The devices neither contain high processing power nor large onboard memory capacities. In contrast, the data rates are in the least range, in kbps to be specific, and battery power is expected to be maintained for several years. Apart from the intended purpose of producing insights using sensor data, security must be validated in this enormous data distribution in real-time. Traditional cryptographic algorithms are computationally demanding for this ecosystem. Consequently, lightweight cipher inventions continue to be introduced in academia in recent years. However, confirmation of sufficient security is still open to question.

Cryptanalysis is as essential as cryptography in performance validation of a cipher in cryptology. In that context, side-channel attack resilience plays a huge role in practical cryptanalysis. Therefore, thorough attention must be kept continuously on these emerging ciphers as the matter is quite critical with short encryption key sizes. In addition, side-channel vulnerabilities are very diverse from invasive to non-invasive and white-box to black-box attack types, *i.e., energy drainage, thermal radiation, optical, EM emission, fault injection, etc*. Yet the highest contribution is seen in power analysis. There are few studies available for EMA as mentioned in section 1. As a result, our analysis covers firmware robustness of PRESENT block cipher against CEMA. No CEMA study regarding PRESENT seems yet available in the literature. SEMA and SFEMA were also covered prior to correlation attack modelling.

Our attack model:

- Uses the maximum number of 256 EM waveforms for 256 different plaintexts.
- Performs a white-box non-invasive attack.
- Reduces noise inference by taking averaged waveforms.
- Produced noisy results.
- Was able to guess one Byte of the encryption key correctly in a random position.

What is more, the identified limitations of the model are:

- A total number of 256 traces may not be enough to cause a sharp output.
- Ambient noise was not affected by aluminium foil coverage.
- Reduced trace data points when the waveform was trimmed to locating the S-box functional area may have reduced the accuracy of the final outcome.

In conclusion, the current analysis outcomes indicate that:

- There are no significant fluctuations in the EM emanation of the Arduino UNO in accordance with the cryptographic operations of PRESENT.
- Encryption key leakage tends to occur in peaks rather than troughs in the resultant correlation graphs.

This work is still on-going, and we expect to optimise our model by:

- Increasing the number of EM trace collections by more than a thousand to enhance accuracy.
- Finding ways to increase the number of data points in the trimmed waveforms.
- Using a Faraday box made of Faraday fabrics for the Arduino UNO placement to reduce ambient noise.
- Applying bandpass filters for the identified frequency leakage range.
- Considering both voltage as well as electric current response in data analytics.
- Comparing the performance on various hardware types if possible, *i.e., FPGA*

Finally, this study further discusses the practical possibilities in successful EM attacks once the encryption runs over all 31 rounds in a commercially-ready manufactured item where accessibility to internal parameters is unavailable.

ACKNOWLEDGEMENTS

We would like to thank Dr. Peiter Robyns, the presenter of [25] for sharing his experience in electromagnetic attack modelling with us throughout our journey.

## AUTHORS

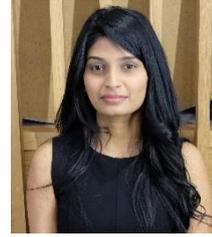

Nilupulee A. Gunathilake is a doctoral student of Blockpass ID Lab in the School of Computing at Edinburgh Napier University, UK. She is a multi-disciplinary researcher who is currently engaged in lightweight cryptography for IoT devices. Her mainstream towards the PhD completion targets side-channel analysis of lightweight ciphers. Nilupulee had previously involved in research in communication engineering related areas; free-space optics (FSO), vehicular communication and microwave propagation. She graduated in MSc (Smart Networks) at the University of the West of Scotland, UK in 2017. Nilupulee obtained a PgDip (Telecommunication and Electronic Engineering), Sheffield Hallam University, UK in 2016 as well as her first degree, BEng (Hons) in Electronics and Communications Engineering, University of Wolverhampton, UK in 2013. She also holds professional qualifications in project management and international spoken English language ESOL. Nilupulee has been involved in academia for over five years and is the first author of five international conference papers. She was a Lecturer in the Faculty of Engineering at Sri Lanka Institute of Information Technology (SLIIT) in 2018 and an Assistant Lecturer (Grade 1) at International College of Business and Technology (ICBT), Sri Lanka in 2015. She worked as a Research Assistant in the Department of Electronic and Telecommunication Engineering at the University of Moratuwa, Sri Lanka in 2016. Nilupulee was offered a PhD scholarship by Edinburgh Napier University in 2018, the Dean's Scholarship by the University of the West of Scotland in 2016 and research grants by the University of Moratuwa in 2015. She was awarded "the University Court Medal" for the best academic excellence of the MSc program.

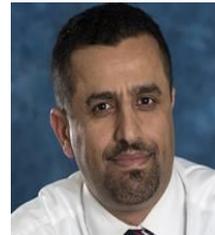

Ahmed Al-Dubai is currently Professor in the School of Computing, Edinburgh Napier University, where he leads the IoT and Networked Systems Research Group. He is also the Director of the Postgraduate Research Degrees Programme. Ahmed was awarded his PhD from the Department of Computing Science, University of Glasgow in 2004. He leads interdisciplinary research and initiatives on the area of Group Communications, High-performance Networks, Internet of Things, Future Networks, E-Health, Smart Cities and Security. His research has been supported by the EU, Universities UK and the Royal Society, Carnegie Trust, EPSRC and Scottish Funding Council. His findings have been published widely in top tier journals including different IEEE Transactions journals, and in prestigious international conferences including IEEE IPDPS, IEEE ICC, IEEE GLOBECOM, IEEE WCNC, ICPP and IEEE IPCCC. Ahmed has been the recipient of several international academic awards and recognition including Best Papers Awards at IEEE IUCC 2015, ACM MoMM 2013 and ACM SAC 2002. He has been regularly invited to give keynote and plenary speeches at several international conferences. He has been a member of several editorial boards of scholarly journals. Ahmed been the Guest Editor for over 25 special issues in scholarly journals and the Chair/Co-Chair of over 30 International Conferences and workshops. He is a Fellow of the British Higher Education Academy and a Senior Member of the IEEE.

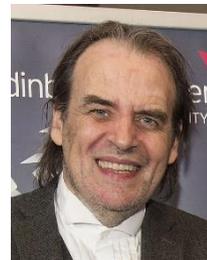

William (Bill) J. Buchanan OBE is a Professor in the School of Computing at Edinburgh Napier University, and a Fellow of the BCS and Principal Fellow of the HEA. He was appointed an Officer of the Order of the British Empire (OBE) in the 2017 Birthday Honours for services to cybersecurity. Bill lives and works in Edinburgh, and is a believer in fairness, justice, and freedom. His social media tagline reflects his strong belief in changing the world for the better: "A Serial Innovator. An Old World Breaker. A New World Creator". He also has a strong belief in the power of education, and in supporting innovation from every angle. Bill currently leads the Blockpass ID Lab and the Centre for Cybersystems and Cryptography. He works in the areas of blockchain, cryptography, trust and digital identity. He has one of the most extensive cryptography sites in the World (asecuritysite.com), and is involved in many areas of novel research and teaching. Bill has published over 30 academic books, and over 300 academic research papers. Along with this, Bill's work has led to many areas of impact, including three highly successful spin-out companies (Zonefox, Symphonic Software and Cyan Forensics), along with awards for excellence in knowledge transfer, and for teaching. Bill recently received an "Outstanding

Contribution to Knowledge Exchange" award, and was included in the FutureScot "Top 50 Scottish Tech People Who Are Changing The World".

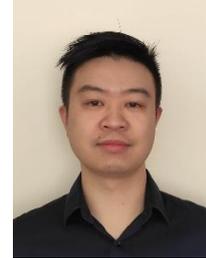

Owen Lo is a Research Fellow (PhD) at Edinburgh Napier University. He obtained a BEng (Hons) in Computer Networks and Distributed Systems at Edinburgh Napier University in 2010 before continuing to complete a PhD on the topic of e-Health in 2015 at the same institute. Awards received by Owen include: Young Software Engineer of the Year Award (Lumison Prize) in 2010, Team Prize Raytheon Cyber Challenge Award in 2011 and Student of the Year ENU Award in 2012. During his PhD, Owen contributed to the Data Capture and Auto Identification Reference (DACAR) Project – a project funded in part by EPSRC and TSB – which aimed to create a secure cloud-based information sharing platform for patient data in healthcare environments. During his time as a researcher at Edinburgh Napier University, Owen has also contributed to the success of two spin-out of two companies: Symphonic Software and Cyan Forensics. For Symphonic Software, Owen helped develop an information-sharing engine used for the secure and trusted sharing of information between different sectors including finance, healthcare, social care and law enforcement. His work on Cyan Forensics included the development of a fully-featured contraband detection software used to rapidly determine if an individual is suspected of storing illegal data on a computer. The contraband detection software was designed specifically to be used by digital forensics experts within the law enforcement sector. Owen's research interests include side-channel analysis, cryptography and computer security. Currently, he is working on a third university spin-out, MemCrypt, which uses novel techniques to combat ransomware.